\documentclass[a4paper]{spie}  
\usepackage{amsmath,amsfonts,amssymb}

\usepackage[utf8]{inputenc}   
\usepackage[english]{babel}   
\usepackage[usenames,dvipsnames]{xcolor} 
\usepackage[colorlinks=true, allcolors=blue, pdftex]{hyperref}
\usepackage{aas_macros}
\usepackage[free-standing-units]{siunitx}
\usepackage{paralist}
\usepackage{graphicx}
\usepackage{float}
\DeclareSIUnit\pixel{px}

\title{Cerberus: A three-headed instrument \\ for the OARPAF telescope}

  \author[a,*]{Lorenzo~Cabona}
  \author[b]{Davide~Ricci}
  \author[c]{Anna~Marini}
  \author[c]{Matteo~Santostefano}
  \author[a]{Matteo~Aliverti}
  \author[d]{Andrea~La~Camera}
  \author[a]{Chiara~Righi}
  \author[c,e]{Silvano~Tosi}

  \affil[a]{INAF-Osservatorio Astronomico di Brera, Via E. Bianchi 46,
    23807, Merate (LC), Italy.}
  
  \affil[b]{INAF-Osservatorio Astronomico di Padova, Vicolo
    dell'Osservatorio 5, 35122 Padova, Italy.}

  \affil[c]{Università degli Studi di Genova, DIFI Dipartimento di
    Fisica, Via Dodecaneso 33, 16146, Genova, Italy.}
  
  \affil[d]{Teiga Srls, Viale Brigate Partigiane 16, 16129,
    Genova, Italy}

  \affil[e]{INFN-Sezione di Genova, Via Dodecaneso 33, 16146 Genova,
    Italy.}

\authorinfo{$^*$Contact information: lorenzo.cabona@inaf.it
}

\begin{document}
\maketitle

\begin{abstract}
  We present the preliminary design of Cerberus, a new scientific
  instrument for the alt-az, $80\centi\meter$ OARPAF telescope in the
  Ligurian mountains above Genoa, Italy.  Cerberus will provide three
  focal stations at the Nasmyth focus, allowing: imaging and
  photometry with standard Johnson-Cousins $UBVRI$+$H\alpha$+Free
  filters, an on-axis guiding camera, and a tip-tilt lens for image
  stabilization up to $10\hertz$; long slit spectroscopy at
  $R\sim5900$ thanks to a LHIRES III spectrograph provided with a
  $1200 l/\milli\meter$ grism; échelle spectroscopy at $R\sim 9300$
  using a FLECHAS spectrograph with optical fiber.
\end{abstract}




\section{Introduction}
\label{sec:intro}

To maximize the scientific return and the quality of the results obtained by the Regional Astronomical Observatory of the Antola Park (OARPAF),  which is in the process of completing the remote control of all its equipment, it was decided to design a new instrument called Cerberus.

Cerberus is an instrument capable of provide three focal stations at the scientific Nasmyth focus of the telescope, and it is composed by an imager, a long slit spectrograph, and an échelle spectrograph.

In Sect.~\ref{sec:obs} we present in general the observatory and its status.  In Sect.~\ref{sec:inst} we show the detectors and the devices that will be used to build the instrument, while the design of Cerberus will be shown in detail in Sect.~\ref{sec:design}.
In Sect.~\ref{sec:aox} we discuss the opportunity to introduce a tip-tilt corrector for photometry in the optical path, and in Sect.~\ref{sec:net} we give an overview of the control network of the instrument in the framework of the observatory network.  Conclusions are treated in Sect.~\ref{sec:conc}.

\section{The Observatory}
\label{sec:obs}

\begin{figure} [t]
  \centering
  \includegraphics[width=0.42\textwidth]{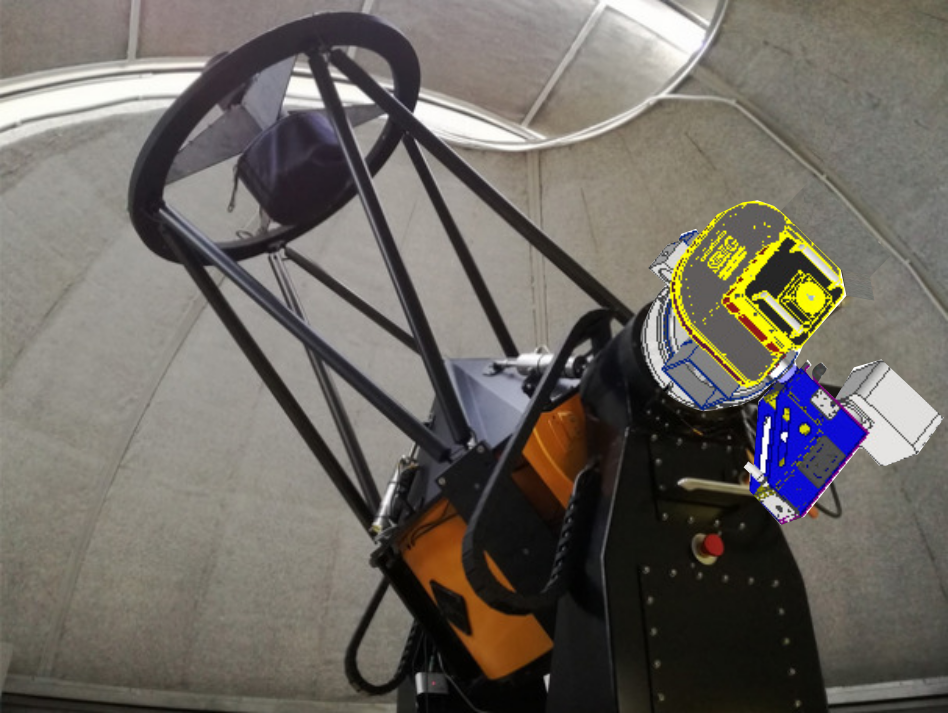} 
  \includegraphics[width=0.57\textwidth]{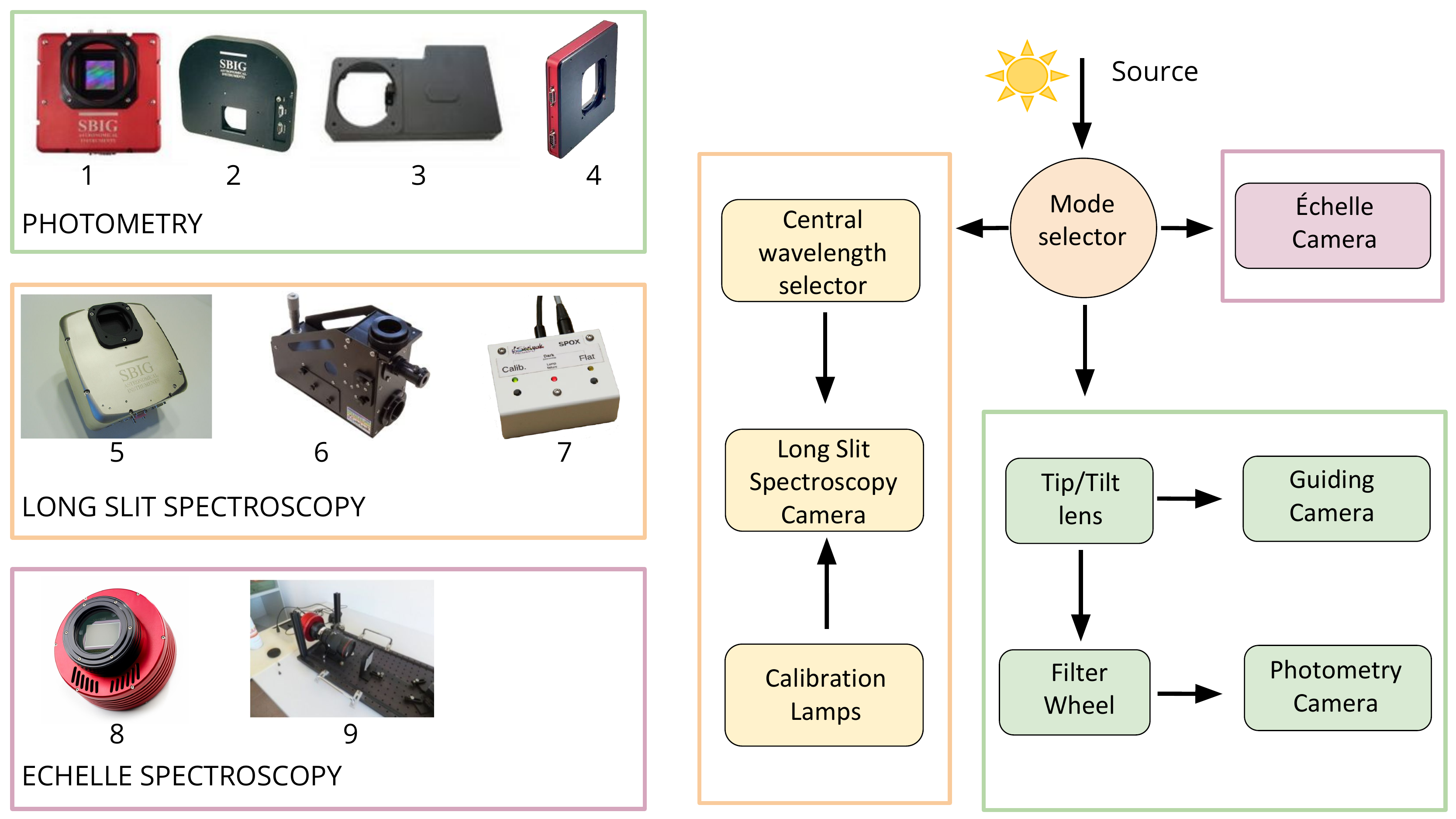}
  \caption{ Left: The OARPAF telescope with a preliminary design of
    Cerberus superposed at the Nasmyth focus. Right: Schematic view of
    Cerberus and references to detectors, accessories and
    spectrographs. }
\end{figure}

OARPAF, inaugurated in 2011\cite{2012ASInC...7....7F}, is situated in the territory of Comune di Fascia, in the Ligurian Apennines at an altitude of about 1480 m above sea level.
  This observatory is one of the few and one of the largest optical telescopes in Italy available in a public facility which allow regular public visits. The observatory on the ground floor of the structure also features a fully equipped conference room, a planetarium, a control room, a library and two guest rooms for scientists.
Currently the internet connection is guaranteed by a private radio link, but it is expected in the near future to receive the cabling via fiber optics, so as to minimize the possibility of failures and increase the data bandwidth.
The telescope installed at OARPAF is an Astelco T0800-01 (see Fig.~1~left), an F8, $80\centi\meter$ diameter Cassegrain telescope with an corrected unvignetted field of view of $45\arcminute$ and two active Nasmyth foci.

The first one, provided with a field derotator, is dedicated to scientific observations, and is shown in the top figure; the second one is dedicated to ocular observations by amateurs.

The scientific Nasmyth focus is currently equipped with an air-cooled \texttt{SBIG STL 11000 M} camera, provided with an internal standard filter wheel and Johnson-Cousins $UBVRI$ filters. The camera is power-connected to an electric socket placed in the fork of the telescope, and USB-connected to a commercial pc inside the fork.  The buffer PC in the fork mount has installed the libsbigudrv library and a \texttt{node.js} software developed by Pierre Sprimont and Davide Ricci, allowing a connection and control via local network using a web browser.

Currently, the telescope is used for imaging/photometry by using this camera.  Our aim is to substitute the current camera with a multi-functional instrument with superior capabilities, providing imaging/photometry, long slit spectroscopy, and a third position for other instrumentation, that will be initially implemented for échelle spectroscopy using the FLECHAS spectrograph connected to an optical fiber.

\section{Instrumentation}
\label{sec:inst}

Also using the funds obtained by the Department of Physics of the University of Genova, thanks to the project named ``Departments of Excellence'', of Italian Ministry for Education, which has as its objective to launching new research lines in astrophysics, as well as an augmented didactic offer in the field, new material has recently been procured for contribute to the full remotization of the OARPAF and the development of the scientific station named Cerberus.
In particular the following instrumentation will be used by this multi-purpose instrument (see Fig.~1~right):

\begin{itemize}
  
\item For imaging/photometry: a SBIG STX-16801 camera (Fig.~$1_1$) with filter
  wheel (Fig.~$1_2$).  In front of the filter wheel an embedded guiding camera (Fig.~$1_3$) and
  a tip-tilt lens (Fig.~$1_4$) are placed.  The following filters will be
  available: $UBVRI$ standard Johnson-Cousins + $H\alpha$ + one
  position without filters.
  
\item For long slit spectroscopy: a LHIRES III spectrograph (Fig.~$1_6$) with two
  gratings available: $2400l/\milli\meter$ and $1200l/\milli\meter$.
  We initially foresee to use the latter grism. The spectrograph is
  provided with a micrometer screw for selecting the central
  wavelength and an internal doublet screw for fine focusing.
  Initially, we foresee to reuse the already owned camera which is
  currently used for science, i.e. the SBIG STL-11000M (Fig.~$1_5$).  The Cerberus
  design avoids to resort to the use of one guiding camera on the
  spectrograph, and uses instead the guiding camera on the
  imaging/photometry path.
  
\item We plan to use the third arm for échelle spectroscopy using the
  optical fiber échelle spectrograph called FLECHAS\cite{flechas} (Fig.~$1_9$)
  already available for the scientific use of the observatory and the
  ATIK11000M camera (Fig.~$1_8$).  This two instruments are already available for
  scientific use but which currently requires the dismount of the
  instrumentation for imaging/photometry.
  
\end{itemize}

\section{Cerberus design}
\label{sec:design}


\begin{figure} [t]
  \centering
  \includegraphics[width=1.00\textwidth]{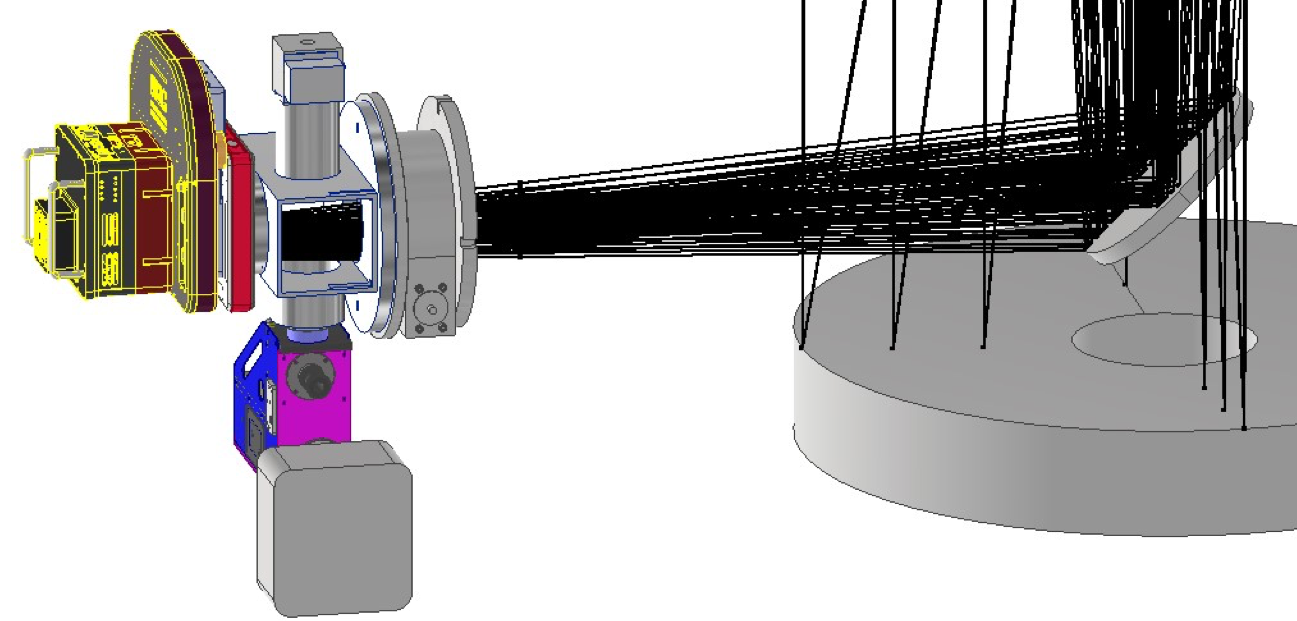}
 \caption{Preliminary opto-mechanical design of Cerberus.}
\end{figure}

The instrument will be located at one of the two Nasmyth interfaces of the telescope, in order to provide three focal stations via a mode selector (see Fig.~2).

We stress here that the mode selector is engineered so that all three configurations will be able to use the embedded guiding camera in the imaging/photometry path. Therefore, using this approach we will be able to minimize the hardware, reducing the costs, weight and volume of the instrumentation applied to the scientific Nasmyth focus.
From this moment on we will refer to the three setup positions as:
imaging/photometry mode, long slit spectroscopy mode, and échelle spectroscopy mode.

\begin{description}

\item[Imaging/photometry mode:]

  This is the main use mode of the Instrument. The light proceed on a
  straight path from the telescope until a tip-tilt lens able to act
  up to a $10\Hz$ rate.  After this lens, the embedded guiding camera
  is placed just before the $50\milli\meter$ filter wheel.  This
  guiding camera, placed in a way that avoids vignetting, is provided
  with a $0.7\times$ focal reducer, and is used to send offset
  information to the telescopes and to the tip-tilt lens, if
  on. Finally, the light focuses on the $4096\times 4096$,
  $9\micro\meter$ pixels array CCD camera for imaging and photometry
  on a $20\arcminute$ FoV with a resolution of
  $0.29\arcsecond/\pixel$.

\item[Long slit spectroscopy mode:]

  This mode requires a flat mirror to direct the light on a slit.  A
  part of the light proceeds straight to the guiding camera on the
  imager.
  The part of the light passing through the slit falls on a
  diffraction grating.  The central wavelength is selected by using a
  micrometer screw that tilts the grating. Then, the light focuses on
  a CCD camera.  We foresee a $R \sim 5800$ resolution in the visible
  band ($450$--$750 \nano\meter$), and the possibility to select the
  central wavelength in order to observe a spectral range of
  $\approx140\nano\meter$ at time.

\item[Échelle spectroscopy mode:]

  This mode requires a flat mirror to direct the light on a
  $15\meter$ long optical fiber. A part of the light is required to
  proceed straight to the guiding camera on the imager. The part of
  the light that passes through the fiber is focused on the CCD camera
  on the Échelle spectrograph FLECHAS \cite{flechas}.

\end{description}


The baseline design for Cerberus will also consists of a central flange, which supports and includes the previous instruments, and, in addition:
\begin{itemize}
\item rotary or linear motors will address and focus the light through
  the several optical paths foreseen by the optical design;
\item switches for lamps and other remote controls will provide
  security and calibration;
\item CCD cameras will take science and calibration frames.
\end{itemize}

\section{Tip-tilt corrector}
\label{sec:aox}

\begin{figure}[t]
  \centering
  \includegraphics[width=0.41\textwidth]{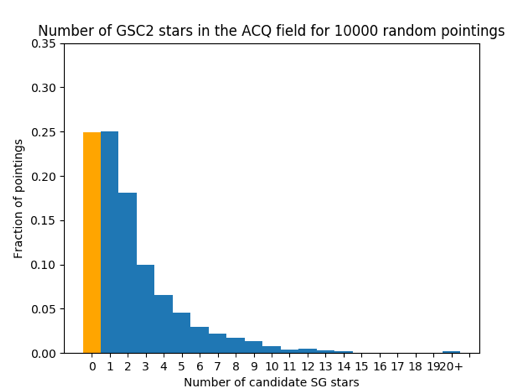} \quad
  \includegraphics[width=0.56\textwidth]{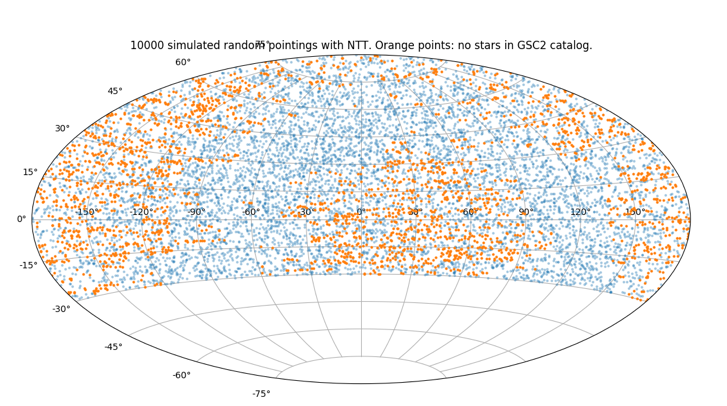}
  \label{fig:aox}
  \caption{ Left: Fraction of pointings with at least $n$ stars
    brighter than $V=13$, based on 10\,000 random pointings. Right:
    Sky position of the 10\,000 random pointings at OARPAF.}
\end{figure}

We evaluated the opportunity to introduce in the optical path the
tip-tilt corrector, called AO-X.  This device consists in a
$10\milli\meter$ BK-7 glass plate that can be software controlled in
tip and tilt ($\pm 2.4\degree$) basing on the centroid position of a
tracking star on the guiding camera.  This closed-loop system can
operate at up to $10\Hz$ and displace the field on the main camera up
to $\pm 144\micro\meter$, i.e. $\pm 16\pixel$ with no significant focal
shift, distortion, rotation, or change in magnification, and with an
accuracy of $0.14\pixel$.

Using the AETC\footnote{Advanced Exposure Time Calculator,
  \url{http://aetc.oapd.inaf.it}} with the Telescope and the guiding
camera parameters, we simulated sets of 100 stars with a $3\arcsecond$
seeing, at several magnitudes and several exposure times.
We fitted each star in each set to check the magnitude limit at which
we are able to retrieve the centroid with an accuracy better than
$0.1\arcsecond$ RMS with a SNR $>10$.
This limit is $V=13$ for an exposure time of $0.1\second$ and a
$3\arcsecond$ seeing.

This information was used to check the fraction of observations that
allows the AO-X to be used in these conditions.  Results on 10\,000
random pointings at OARPAF, querying the GSC-II catalog (see Fig.~3), show that:
\begin{itemize}
\item 25\% of the pointings have no stars brighter than $V=13$;
\item 25\% of the pointings have one star brighter than $V=13$.
\item 50\% of the pointings have more than one star brighter than $V=13$.
\end{itemize}
This suggests that we are able to correct at $\approx 10\Hz$ small
tip-tilt displacements of the field due to telescope tracking error,
or mechanical/atmosphere issues, for a large fraction of the
observations.
This fraction can in principle be more favourable with better seeing
conditions, or if we decide to correct at smaller frequences.

\section{Planned Network Layout}
\label{sec:net}

\begin{figure} [t]
  \centering
  \includegraphics[width=0.9\textwidth]{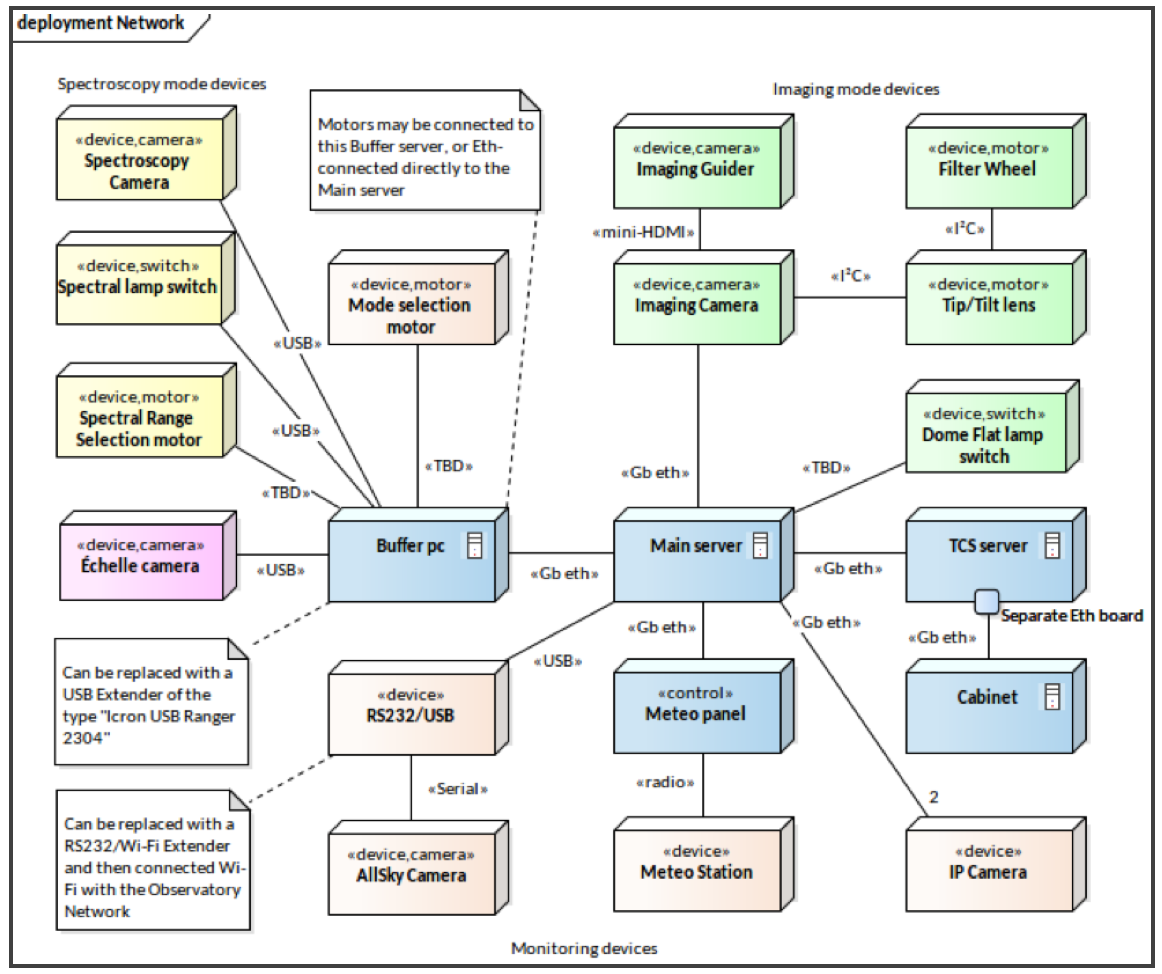}
  \caption{ Network of the components of the OARPAF observatory. }
\end{figure}

Cerberus has three detectors, a guiding camera, and a series of
motorized functions.  All these components must be part of a control
network.

All components relative to the photometry head are ethernet
connected via the imaging Camera directly to a main server at the
lower floor of the telescope, via a ethernet switch installed in the
telescope fork.

The components relative to the spectroscopy heads, as the mode
selector, will be mainly USB devices.  Due to the limited range of the
USB connections, a small buffer pc placed in the telescope fork will
collect these connections.  This buffer pc will be ethernet connected
to the main server. A fallback option consists in a USB extender such
as the ``Icron USB Ranger 2304'', which has been recently tested in
the framework of the SOXS instrument for
ESO-NTT\cite{2018SPIE10707E..1GR}.


This instrument is part of the network of the observatory, which
foresees other components, such as of course the telescope TCS
server, the Meteo station, the IP cameras, the AllSky Camera, and the
remote control of dome flat field lamps, all connected to the main
server.  The entire deployment network can be seen in Fig.~4.

\section{Conclusion}
\label{sec:conc}

The OARPAF observatory can give a valuable contribution in cutting-edge scientific topics, such as the search for exoplanets, the observation of AGNs and the study of asteroids and at the same time be a point of reference for students at the University of Genova who decide to approach experimental Astrophysics and technological development.  Thanks to its great potential, the funding needed for the complete remotization of the facility and allow to improve the scientific instrumentation has been obtained. Therefore, OARPAF is expected to fully operate remotely by 2021.

In order to allow astronomers to remotely use all the scientific instrumentation, the multi-functional instrument named Cerberus has been designed.  Cerberus include an imager, a long slit spectrograph, and an échelle spectrograph.  The imager can also benefit of tip-tilt correction up to $10\Hz$, that will be available at its top frequence for a large part of observed fields.

Once the feasibility study has been completed, we will proceed with the construction of this instrument, which will significantly increase the scientific return of the telescope.


\bibliographystyle{spiebib} 
\bibliography{biblio} 

\end{document}